\documentstyle[11pt,paspconf]{article}

\def\fun#1#2{\lower3.6pt\vbox{\baselineskip0pt\lineskip.9pt
  \ialign{$\mathsurround=0pt#1\hfil##\hfil$\crcr#2\crcr\sim\crcr}}}
\def\lap{\mathrel{\mathpalette\fun <}}
\def\gap{\mathrel{\mathpalette\fun >}}
\def\mhole{M_{\bullet}}
\def\msun{M_{\odot}}

\begin{document}

\title{Black Holes and Galaxy Evolution}

\author{David Merritt}
\affil{Department of Physics and Astronomy, Rutgers University}

\begin{abstract}
Supermassive black holes appear to be generic components of 
galactic nuclei.
Following their formation in the early universe, black holes 
should often find themselves in bound pairs as a consequence of 
galaxy mergers.
The greatest uncertainty in estimating the coalescence time of black hole
binaries is the degree to which a binary wanders about the 
center of the galactic potential. 
A simple model for binary decay is 
presented which qualitatively reproduces the evolution 
observed in the $N$-body simulations.
The model predicts binary coalescence times that are never 
less than several billion years.
Mass ejection by a decaying black hole binary should 
substantially lower the density of its host nucleus.
The weak density cusps of bright ellipticals may be explained in 
this way if it is assumed that these galaxies formed from nearly 
gas-free mergers.

\end{abstract}


\section{Introduction}

The case for supermassive black holes (BHs) in galactic nuclei is
now very strong, and in a few galaxies almost irrefutable.
The evidence is most compelling in galaxies where the
kinematics of stars or gas can be traced into sub-parsec scales;
examples are the Milky Way (Genzel et al. 1997; Ghez et al.
1998), where stellar proper motions can be measured at distances
of $\lap 0.02$ pc, and NGC 4258 (Miyoshi et al. 1995), where
the rotation curve of H$_2$O maser sources obeys Kepler's law
into $\sim 0.13$ pc.
For a larger sample of galaxies, kinematical data extending into
$\sim 10$ or $\sim 100$ pc provide suggestive though not
yet irrefutable evidence of supermassive compact objects
(Kormendy \& Richstone 1995).

Although the masses of the detected BHs comprise on average only
$\sim 0.3\%$ of the mass of their host spheroids (Ho 1998),
there is a growing body of work suggesting that the dynamical 
influence of a supermassive BH can extend far beyond the nucleus.
Furthermore the formation and growth of BHs may be intimately
connected with the evolution of galaxies on larger scales.
For instance, mergers between galaxies containing nuclear BHs
would produce supermassive BH
binaries which would eventually coalesce via the emission of
gravitational radiation (Begelman, Blandford \& Rees 1980).
The formation and decay of these binaries 
may be relevant to a wide range of phenomena, from
the wiggling of radio jets (Kaastra \& Roos 1992)
to the destruction of stellar density cusps (Ebisuzaki et al. 1991).

This review discusses the time scale for coalescence of 
supermassive BH binaries (\S 2) and the dynamical effect of the BHs on 
stellar nuclei (\S3) and the large-scale structure of galaxies (\S4).

\section{Supermassive Black Hole Binaries}

Supermassive BHs appear to be ubiquitous components of galactic nuclei.
Since galaxies often merge, one would expect to form BH binaries at a rate
that is roughly equal to the galaxy merger rate.
The formation and coalescence of a BH binary is believed to take
place via three, fairly distinct stages (Begelman, Blandford \& Rees 1980):

1. Two parent galaxies interact; the BHs -- surrounded by their
dense star clusters -- sink to the center of the common potential
well via dynamical friction, forming a binary.

2. The BH binary shrinks by ejecting stars from the nucleus via
three-body interactions.

3. When the binary separation has fallen to the value at which
gravitational radiation becomes efficient, the BHs coalesce.

\noindent
In the absence of gas, the rates of the first two processes are 
in principle possible to calculate via $N$-body simulations.
However the problem is a difficult one to simulate due to the wide
range of length and time scales; none of the published simulations have
succeeded in following the decay to the point where gravitational radiation 
would become important.
Here we derive a simple analytical model for the decay of BH binaries
which reproduces the results of the $N$-body simulations, then 
use it to estimate the coalescence time.

Consider a binary with total mass 
$M_{12}=m_1+m_2$, $m_1\geq m_2$ and semimajor axis 
$a(t)$.
We assume that the background galaxy is initially a
singular isothermal sphere, with density and mass profiles
\begin{equation}
\rho_0(r) = {\sigma^2\over 2\pi G r^2}, \ \ \ \ M_0(r) = {2\sigma^2r\over 
G};
\end{equation}
$\sigma$ is the one-dimensional stellar velocity dispersion.
The binary becomes ``hard'' when $a$ falls below 
$\sim a_h=Gm_2/4\sigma^2$ (Quinlan 1996); 
subsequent evolution is driven by the 
capture and ejection of stars that interact with the binary.
The hardening rate
\begin{equation}
H = {\sigma\over G\rho}{d\over dt}\left({1\over a}\right)
\end{equation}
depends only weakly on $m_1/m_2$ and on $a$ for $a\lap a_h$;
in the limit $a\ll a_h$, $H\approx 16$ 
(Hills 1992; Mikkola \& Valtonen 1992; Quinlan 1996).
Mass ejection occurs at a rate
\begin{equation}
J = {1\over M_{12}}{dM_{ej}\over d\ln(1/a)}
\end{equation}
where $J\approx 1$ is again nearly independent of ($m_1/m_2$, $a$) 
for $a\ll a_h$ (Quinlan 1996).
Thus 
\begin{equation}
M_{ej} \approx J M_{12} \ln(a_h/a).
\end{equation}

If the binary remained fixed at the center of a spherical potential, 
it would quickly eject all stars on orbits with pericenters less 
than $\sim a$.
The local density would fall to zero and the binary would cease 
to harden.
We assume instead that mass removal produces a 
constant-density core of stars out to some radius $r_c(t)$, where
\begin{equation}
M_{ej} = M_0(r_c) - {4\pi\over 3}\rho_0(r_c)r_c^3 = {4\over 
3}{\sigma^2 r_c\over G}.
\end{equation}
In order to achieve this, the binary must interact with stars 
at radii $r\lap r_c\gg a$.
This may seem an optimistic assumption, but it turns out that 
something very similar happens in the $N$-body simulations, 
as discussed below.

\begin{figure*}
\includegraphics{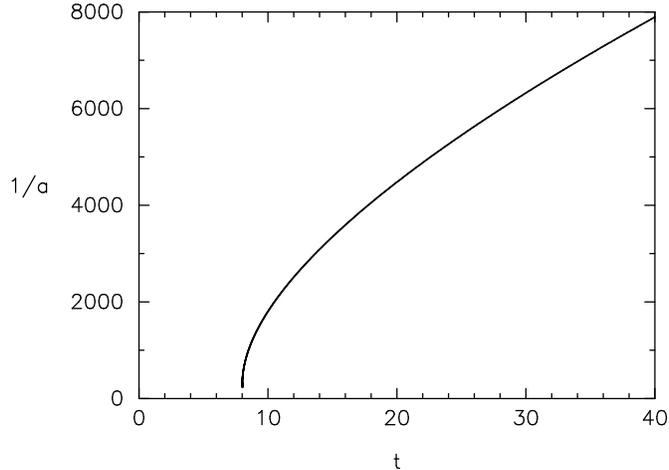}
\vspace{6cm}
\caption{Dependence of binary semimajor axis $a$ on time for the analytic
model of eq. (7).
Units are discussed in the text.}
\end{figure*}

The binary separation rapidly falls below $r_c$ and 
subsequent evolution occurs against an evolving background 
density $\rho(t) = \sigma^2/2\pi G r_c^2(t)$.
Combining equations (2), (4) and (5), we find for the hardening 
rate
\begin{equation}
{d(a_h/a)\over d(t/t_0)} = {1\over\ln^2(a_h/a)},\ \ \ \ \ 
t_0={9\pi J^2\over H}{\left(M_{12}\over 2m_2\right)}
{\left(GM_{12}\over\sigma^3\right)}.
\end{equation}
The logarithm represents the decrease in the hardening rate due 
to the declining stellar density.
Integrating with respect to time,
\begin{equation}
{t-t_h\over t_0} = {a_h\over a}\left[\ln^2\left({a_h\over 
a}\right) - 
\ln\left({a_h\over a}\right)^2 + 2\left(1-{a\over 
a_h}\right)\right]
\end{equation}
where $a(t_h)=a_h$.
The functions $t(a)$ and $a(t)$ may be approximated as 
\begin{equation}
{t-t_h\over t_0} \approx 0.6\left({a_h\over a}\right)\ln^2
\left({a_h\over a}\right),\ \ \ \ \ \ 
{a_h\over a} \approx 
{4(t-t_h)/t_0\over\ln^2\left[(t-t_h)/t_0\right]}
\end{equation}
for $t-t_h\gap $ a few $\times\ t_0$.

We may compare the predictions of this simple model to the results of 
$N$-body simulations of BH binary decay.
Quinlan \& Hernquist (1997) (QH) followed the evolution of a BH 
binary in a galaxy with an initial $\rho\propto r^{-2}$ 
density cusp.
Expressed in their units, the scaling parameters in eq. (7) are
$a_h=1/200$, $t_0\approx 1/10$ and $t_h\approx 8$.
Fig. 1 is a plot of eq. (7) scaled to these units.
The predicted time dependence $a(t)$ is qualitatively very
similar to that found by QH (their Fig. 2).
However, those authors found that the rate of binary decay depended 
systematically on the number $N$ of particles used to represent the 
galaxy: larger values for $N$ gave slower evolution.
For $N=6250$, the smallest number considered by them,
Fig. 1 overestimates the evolution rate by a factor of 
only $\sim 2$, remarkably close given the approximations made.
For $N=2.5\times 10^4$ and $10^5$, 
the adjustment factors are $\sim 3$ and $\sim 5$ respectively.

QH attributed the $N$-dependence of the hardening rate to
wandering of the binary about the center of the galaxy 
potential.
In a core of density $\rho$, an rms velocity $v_b$ of the 
binary's center of mass will produce a wandering 
with amplitude $r_w \sim v_b/(G\rho)^{1/2}$.
If the source of the binary's motion were elastic encounters with 
stars of mass $m_*$, equipartition arguments would give 
$v_b/\sigma\approx (m_*/M_{12})^{1/2}\approx 10^{-4}$.
However the binary converts some of its binding energy into 
bulk motion when it ejects stars, with momenta $\sim 
m_*(a_h/a)^{1/2}\sigma$.
This ``super-elastic scattering'' allows the binary to sample the 
stellar density over a much wider range of radii than if it remained 
stationary;
QH in fact verified that fixing the binary to the center of the 
potential caused the hardening to cease after the binary had 
ejected most of the stars within $r=a_h$.

If the background density within the wandering radius $r_w$ were
to fall, the gravitational restoring force acting on the binary
would also drop and the wandering amplitude would increase, 
roughly in proportion to $\rho^{-1/2}$.
This feedback mechanism should tend to inhibit the formation of a central 
``hole'' in the stellar density (Mikkola \& Valtonen 1992).
The crude model derived above mimics this process by forcing the 
background density to remain constant for $r< r_c$;
the model could be improved by incorporating the wandering radius 
explicitly.

In the $N$-body simulations, the perturbations that the binary 
experiences from the ``stars'' become less noisy as $N$ is 
increased, implying a reduced wandering radius and a lower 
hardening rate.
Makino (1997) found that the wandering amplitude 
varied roughly as $N^{-1/2}$ for $N<3\times 10^5$.
However QH found that the hardening rate did not change appreciably 
when $N$ was increased from $10^5$ to $2\times 10^5$.
They argued that $r_w$ might reach a limiting amplitude for 
large $N$ due to the feedback mechanism discussed above.
This hypothesis is important to check; however doing so will require 
simulations with very large particle numbers.


The binary will rapidly coalesce when the time scale for 
emission of gravitational radiation
\begin{equation}
t_{gr} = {5\over 256}{c^5a^4\over G^3 m_1m_2M_{12}}
\end{equation}
(Peters 1964, for a circular binary) equals the hardening time $|a/\dot a|$.
Using eq. (6), this occurs when $a=a_{gr}$ where
\begin{equation}
{\left(a_{gr}/a_h\right)^5\over \ln^2\left(a_{gr}/ 
a_h\right)} = {9\pi\times 16^5 J^2\over 20H} 
\left({m_1\over m_2}\right) \left({M_{12}\over 
2m_2}\right)^2\left({\sigma\over c}\right)^5
\end{equation}
or
\begin{equation}
a_{gr}/a_h \approx A |\ln A |^{0.4},\ \ \ \ \ \ 
 A = 9.85\left({m_1\over m_2}\right)^{0.2} \left({M_{12}\over 
2m_2}\right)^{0.4} \left({\sigma\over c}\right).
\end{equation}
For $\sigma/c = 300/(3\times 10^5) \approx 0.001$, $A\approx 0.01$ and
\begin{equation}
a_{gr}/a_h\approx 0.018,
\end{equation}
 i.e. the binary must shrink by a factor of 
$\sim 50$ beyond the hardening radius for coalescence to occur.
The simulations of QH followed the decay only over a 
factor of $\sim 12$ beyond $a_h$.

\begin{figure*}
\includegraphics{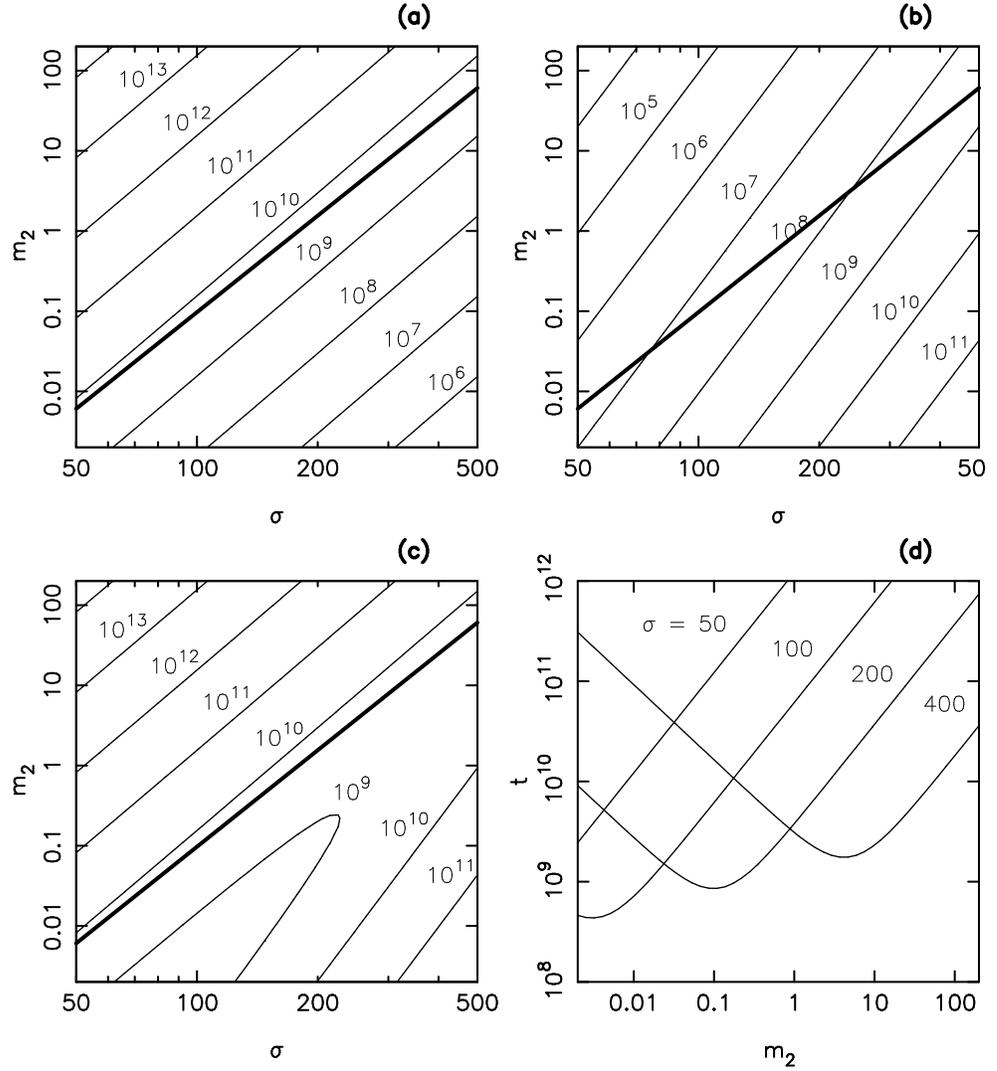}
\vspace{15cm}
\caption{Decay time in years for supermassive BH binaries in singular 
isothermal sphere galaxies. 
$m_2$ is the mass of the smaller BH in $10^8 \msun$; 
$\sigma$ is the one-dimensional velocity dispersion of the larger 
galaxy in km s$^{-1}$.
(a) Time $\Delta t_1=t_{gr}-t_h$ from formation of a hard binary to 
coalescence by emission of gravitational radiation.
The binary mass ratio is assumed to be $m_2/m_1=1:2$.
The solid line is the ``Faber-Jackson relation'' for BHs (see 
text).
(b) Orbital decay time $\Delta t_2$ of the smaller galaxy (containing 
the BH of mass $m_2$) due to dynamical friction against the larger 
galaxy (of velocity dispersion $\sigma$).
(c) Total evolution time $T = \Delta t_1 + \Delta t_2$.
(d) Evolution time as a function of $m_2$ for $\sigma = (50, 100, 
200, 400)$ km s$^{-1}$.}
\end{figure*}

We can use the analytic expressions derived here to extend the 
$N$-body simulations until a time corresponding to coalescence; 
the correction factor of 5 mentioned above is applied to $(a/a_h)$ 
in order to reproduce the hardening rate observed in the largest-$N$ 
simulations. 
Fig. 2a shows the time from  formation of a hard binary until 
coalescence, $\Delta t_1$, as a function of 
$\sigma$ and $m_2$ for a binary with $m_2/m_1=1/2$.
An approximate expression is
\begin{equation}
\Delta t_1 \approx 8t_0 A^{-1}|\ln A|^{8/5};
\end{equation}
if we further approximate the logarithm as a constant, $|\ln A|\approx |\ln 
0.01| \approx 4.61$, then
\begin{equation}
\Delta t_1 \approx 1.4\times 10^{10} {\rm yr} \left({m_2\over m_1}\right)^{0.2} 
\left({M_{12}\over 2m_2}\right)^{0.6} \left({M_{12}\over 
10^9M_{\odot}}\right) \left({\sigma\over 200 {\rm \ km\ s}^{-1}}\right)^{-4}.
\end{equation}
While eq. (14) is an accurate extrapolation of the evolution seen 
in the $N$-body experiments, it should probably be interpreted as 
a lower limit to the coalescence time expected in a real 
spherical galaxy due to the likely overestimate of the wandering 
radius in the simulations.
However the dependence of $\Delta t_1$ 
on $m_2$ and $\sigma$ is probably more robust.
Since for real galaxies $L\sim \sigma^4$ and $M_{12}\sim L$, eq. 
(14) predicts a coalescence time that is almost independent of 
$\sigma$ and $M_{12}$.
The solid line in Fig. 2a is the ``Faber-Jackson 
law'' for BHs,
\begin{equation}
{\mhole\over 10^8M_{\odot}} \approx 3.1 
\left({\sigma\over 200{\rm \ km\ s^{-1}}}\right)^4;
\end{equation}
this relation is consistent with three galaxies whose BH masses
are well-determined: 
M84 (Bower et al. 1998), 
M87 (Macchetto et al. 1997) and 
NGC 4258 (Miyoshi et al. 1995).
Fig. 2a suggests a typical coalescence time of several billion 
years.

Very low mass binaries would evolve more rapidly (eq. 14), 
but another factor would delay their coalescence: the necessity of 
the two BHs finding their way to the potential center following a 
galaxy merger.
Governato et al. (1994) carried out merger 
simulations of galaxies containing BHs; 
they concluded that inspiral times could be 
very long if the infalling galaxy had a sufficiently 
low mean density to be tidally disrupted.
A more likely scenario is that the infalling BH retains some of 
the mass of its host galaxy.
We can estimate the inspiral time 
by assuming that the smaller galaxy is tidally limited by 
the larger galaxy as its orbit decays. 
Following Merritt (1984),
the tidal radius is located at the point
where the effective potential 
(gravitational plus centrifugal) has a saddle point.
Approximating both galaxies as singular isothermal spheres
and assuming that the smaller galaxy (of mass $m_g$) is 
on a circular orbit of radius $r$, its tidal radius and
mass become
\begin{equation}
r_g^3 \approx  {Gm_gr^2\over 4\sigma^2}, \ \ \ \ 
m_g \approx {\sigma_g^3r\over 2G\sigma}
\end{equation}
with $\sigma$ and $\sigma_g$ the velocity dispersions of the 
larger and smaller galaxies respectively.
Chandrasekhar's formula then gives for the orbital decay rate
and infall time
\begin{equation}
{dr\over dt} = -0.30{Gm_g\over\sigma r} \ln\Lambda \approx 
-0.151{\sigma^3_g\over\sigma^2}\ln\Lambda, \ \ \ \ 
\Delta t_2 \approx 0.30 {r_e\sigma^2\over\sigma^3_g}
\end{equation}
(Tremaine 1990), 
where $\ln\Lambda\approx \ln(r/0.5r_g)\approx \ln(4\sigma/\sigma_g)\approx 
2$.
Taking $r(0) = r_e\approx 2.6\ {\rm kpc} (\sigma/200 
\ {\rm km\ s}^{-1})^3$ (Valluri \& Merritt 1998), 
the effective radius of the larger 
galaxy, the inspiral time becomes
\begin{equation}
\Delta t_2 \approx 9.6\times 10^7 {\rm yr} \left({\sigma\over 200 
\ {\rm km s}^{-1}}\right)^5\left({m_2\over 
10^8M_{\odot}}\right)^{-3/4}
\end{equation}
where eq. (15) has been used to relate $\sigma_g$ to $m_2$.
This relation is plotted in Fig. 2b and the total time 
$t_{tot} = \Delta t_1 + \Delta t_2$ in Fig. 2c.
Remarkably, Fig. 2c suggests that real BHs have roughly the mass 
that would be required to minimize their total coalescence time.
Nevertheless, this time appears to be very long.

Other mechanisms might accelerate the coalescence.
If the potential of the stellar nucleus is non-axisymmetric, 
orbits will not 
conserve angular momentum and stars with a much wider range of 
energies may interact with the binary.
However it is not clear whether self-consistent triaxiality can
be maintained in a nucleus where the potential is dominated by a BH
(Sridhar \& Touma 1999; Merritt \& Valluri 1999).
Gas if present would also exert a drag force on the binary.
``Loss-cone refilling'' by two-body interactions between stars 
(Frank \& Rees 1976) is almost certainly unimportant due to the long 
relaxation times in nuclei (Valtonen 1996).

If coalescence times are comparable to the mean time between 
galaxy mergers, a third BH would often be introduced  
into a nucleus containing an uncoalesced pair.
The most likely outcome is ejection of two BHs in opposite 
directions with the third remaining at the center
(Valtonen 1996).
This process could substantially reduce the mean ratio of BH mass 
to galaxy mass over time.

\section{Formation and Destruction of Nuclei}

Combining eqs. (4) and (12), the total mass ejected by the BH binary 
between $t_h$ and $t_{gr}$ is $\sim 4 M_{12}$.
The structure of the pre-existing nucleus should therefore be 
disturbed out to a radius where the enclosed 
stellar mass was a few times $M_{12}$.
A reasonable guess for the initial density profile is $\rho\approx
r^{-\gamma}, \gamma\approx 2$, as assumed in the model above; 
adiabatic growth of a BH in a nucleus with a shallower profile generically 
leads to a power-law cusp with index near 2 (Quinlan et al. 1995; 
Merritt \& Quinlan 1998), and this is also the slope observed in 
the faintest ellipticals (Kormendy et al. 1995)
which are the least likely to have 
experienced the sort of core disruption discussed above.

Simply removing all of the stars with energies below some minimum
$E_{min}$ produces a density profile near the BH of
\begin{equation}
\rho(r) = 4\pi\int_{E_{min}}^0 
f(E)\sqrt{2\left[E-\Phi(r)\right]}dE \approx 
C\times f(E_{min})\sqrt{{GM_{12}\over r}} \propto r^{-1/2}
\end{equation}
assuming an isotropic velocity distribution $f(E)$.
Nakano \& Makino (1999a,b) found this to be a good description 
of what happens in $N$-body simulations 
of a single BH spiralling into a pre-existing core.
In the case of a decaying BH binary, 
stars on eccentric orbits are most likely to interact with the 
binary and be removed,
leaving behind a nucleus in which the stellar motions are 
strongly biased toward circular.
The resulting density profile is difficult to calculate.
The $N$-body simulations discussed above (Makino 1997; QH) 
yield central profiles that 
are crudely describable as power laws, with indices 
$0\lap\gamma\lap 1$; 
however these results are probably dependent on the 
degree of BH wandering and hence on $N$.

Which galaxies in the current universe would be expected to 
contain these low-density cores?
Kauffmann \& Haehnelt (1999) present a semi-analytic model for 
the formation of galaxies in a cold-dark-matter 
universe; they assume that supermassive BHs form from gas that is 
driven into the centers of galaxies during mergers.
The predicted ratio of gas mass to stellar mass during the last major 
merger is a steep function of galaxy luminosity in their model;
this ratio drops from $\sim 3$ for $M_v=-18$ to $\sim 0.3$ for 
$M_v=-21$, albeit with considerable scatter (their Fig. 14). 
Gas-rich mergers would be expected to form dense nuclei while 
gas-free mergers should produce low-density cores after the BHs 
coalesce.
Figure 3 shows that the variation of $\gamma$ with 
galaxy luminosity is roughly consistent with Kauffmann \& Haehnelt's 
model.
Faint ellipticals, $M_v\gap -19$, have $\gamma\approx 2$, 
while for brighter galaxies
$\gamma$ decreases with increasing luminosity.
This trend is often described as a dichotomy although the data 
of Fig. 3 suggest a continuous distribution, 
as expected in a model like Kauffmann \& Haehnelt's.

\begin{figure*}
\includegraphics{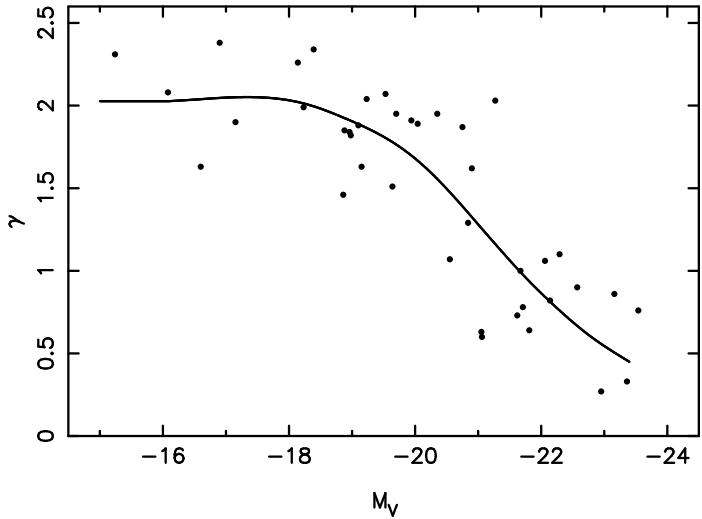}
\vspace{7cm}
\caption{Luminosty density slopes, $\rho\sim r^{-\gamma}$, for 
elliptical galaxies vs. absolute magnitude 
(adapted from Gebhardt et al. 1996).
The solid line is a nonparametric regression fit to $\gamma(M_v)$.}
\end{figure*}

\section{Large-Scale Evolution}
The gravitational influence of a supermassive BH can extend far 
beyond the nucleus in a galaxy that is not axisymmetric
(Gerhard \& Binney 1985).
The mechanism is dynamical chaos induced in the stellar orbits by 
close passages to the BH (Merritt \& Valluri 1996).
Although relatively little $N$-body work has been done on this 
problem, one study (Merritt \& Quinlan 1998) found that nuclear 
point masses can cause initially triaxial galaxies to evolve to 
globally axisymmetric shapes in little more than a crossing time
when the ratio of BH mass to galaxy mass exceeds $\sim 2\%$.
This mass ratio is consistent with the value that induces a 
transition to global stochasticity 
and rapid mixing in the phase space of 
box-like orbits (Valluri \& Merritt 1998).

Mergers of stellar disks produce generically triaxial objects
(Barnes 1996); adding a dissipative component to the simulations 
produces end states that are much more nearly axisymmetric 
(Barnes \& Hernquist 1996).
The evolution in shape occurs rapidly once a few 
percent of the total mass has accumulated in the center 
(Barnes 1999).
The stars in these simulations respond to the ``gas'' only 
insofar as the latter affects the potential; thus the change in 
shape is a purely stellar-dynamical phenomenon, and we would 
expect to see similar evolution even in gas-free mergers if the 
BHs are sufficiently massive.

The typical ratio of BH mass to galaxy mass is $\sim 
0.003$ (Ho 1998), somewhat less than the value required
to induce a rapid transition to axisymmetry.
For $\mhole/M_{gal}=0.003$, Merritt \& Quinlan (1998) 
found a time scale of 
$\sim 10^2$ crossing times for the evolution to axisymmetry.
This exceeds a Hubble time for galaxies with 
luminosities above $M_v\approx -19$ (Valluri \& Merritt 1998); 
hence we might expect bright ellipticals to often retain their 
merger-induced triaxial shapes.
In fact there is evidence for a systematic change in the shape of 
ellipticals at about this magnitude (Tremblay \& Merritt 1996); 
bright ellipticals as a class show evidence for triaxiality, 
while the axis-ratio distribution of faint ellipticals is 
consistent with axisymmetry.

The same orbital evolution that destroys triaxiality also tends 
to produce a smoother phase-space density.
One consequence is that ``boxiness'' in the isodensity contours 
-- a natural consequence of a non-smooth phase space distribution 
(Binney \& Petrou 1985) -- should be destroyed along with 
triaxiality.
This too has been observed in simulations of gaseous mergers; 
the effect is sometimes attributed to ``dissipation'' (e.g. 
Bekki \& Shioya 1997) but it is again almost certainly a 
purely stellar dynamical effect.
A prediction is that triaxiality should correlate with 
boxiness, since a central mass concentration or BH 
should tend to destroy both following a merger.
Evidence for such a correlation has been noted by Kormendy \& 
Bender (1996).

\acknowledgments
I thank M. Haehnelt for useful discussions.

\end{document}